\documentclass[a4paper,11pt]{article}
\pdfoutput=1 
\usepackage{jheppub} 

\usepackage[T1]{fontenc} 
\usepackage[toc,page]{appendix}
\usepackage{empheq}
\usepackage{graphicx}
\usepackage{subcaption}
\usepackage{todonotes}

\newcommand{\cN}{\mathcal{N}}

\newcommand{\nn}{\nonumber}

\newcommand{\vp}{\varphi}

\renewcommand{\(}{\left(}
\renewcommand{\)}{\right)}

\newcommand{\cG}{\mathcal{G}}

\def\K{K{\"a}hler}
\def\be{\begin{equation}}
\def\ee{\end{equation}}
\newcommand{\ba}{\begin{eqnarray}}
\newcommand{\ea}{\end{eqnarray}}

\newcommand{\rf}[1]{(\ref{#1})}

 \title{\rm {\bf \huge {\boldmath {Universality of multi-field $\alpha$-attractors}}}}

\author [a,b]{Ana Ach\'ucarro,}
\author[c]{Renata Kallosh,}
\author[c]{Andrei Linde,}
\author[a,d]{Dong-Gang Wang}
\author[a,d]{and Yvette Welling}
\affiliation[a]{Lorentz Institute for Theoretical Physics, Leiden University, 2333CA Leiden, The Netherlands}
\affiliation[b]{Department of Theoretical Physics, University of the Basque Country, 48080 Bilbao, Spain}
\affiliation[c]{Stanford Institute for Theoretical Physics and Department of Physics, Stanford University, Stanford, CA 94305, USA}
\affiliation[d]{Leiden Observatory, Leiden University, 2300 RA Leiden, The Netherlands}
\emailAdd{achucar@lorentz.leidenuniv.nl}
\emailAdd{kallosh@stanford.edu}
\emailAdd{alinde@stanford.edu}
\emailAdd{wdgang@strw.leidenuniv.nl}
\emailAdd{welling@strw.leidenuniv.nl}

\notoc

\abstract{We study a particular version of the theory of cosmological $\alpha$-attractors with $\alpha=1/3$, in which both the dilaton (inflaton) field and the axion field are light during inflation. The kinetic terms in this theory originate from  maximal $\cN=4$ superconformal symmetry and from  maximal $\cN=8$ supergravity.  We show that because of the underlying hyperbolic geometry of the moduli space in this theory, it exhibits double attractor behavior: their cosmological predictions are stable not only with respect to significant modifications of the dilaton potential, but also with respect to significant modifications of the axion potential:  $n_s\simeq 1-{2\over N}$, $r\simeq {4\over N^2}$.
We also  show that the universality of predictions extends to
other values of $\alpha \lesssim {\cal O}(1)$  with general two-field potentials
that may or may not have an embedding in supergravity.
Our results support the idea that inflation involving multiple, not stabilized, light fields on a hyperbolic manifold
may be compatible with current observational constraints for a broad class of potentials.}

\begin{document}

\maketitle

\newpage

 \tableofcontents{}

\section{Introduction}

UV embeddings of inflation typically contain multiple scalar fields beside the inflaton.
If the additional fields are stabilized, we can integrate them out to find effectively single field inflation. On the other hand, if the additional fields remain light during inflation, we should take into account the full multi-field dynamics. Planck \cite{Ade:2015lrj, Ade:2015ava} puts tight constraints on these inflationary models, therefore we should understand which model-building ingredients are important to ensure compatibility with the data. In particular, both the geometry of field space and the curvature of the inflationary trajectory play a very important role in determining the observables. In this paper we focus on the special role played by hyperbolic geometry.

A notable example are the $\alpha$-attractor models, a relatively simple class of inflationary models that have a single scalar field driving inflation.
In the simplest supergravity embedding of these models, the potential  depends on the complex scalar $Z= \rho\, e^{i\theta} $, where $Z$ belongs to the Poincar\'e disk with $|Z| = \rho < 1$ and the kinetic terms read\footnote{Alternatively,  $3\alpha  {\partial T \partial \bar T\over (T+\bar T)^2} $,  where $T = {1+Z\over  1-Z} $.}
\be  3\alpha {\partial_\mu \bar Z \partial^\mu Z \over (1 - Z \bar Z)^2  } + ...
\label{geometry}
\ee
In many versions of these models, the field $\theta$ is heavy and stabilized at $\theta = 0$,  so that the inflationary trajectory corresponds to the evolution of the single field $\rho$. An important property of these models is that their cosmological predictions are stable with respect to considerable deformations of the choice of the potential of the field $\rho$:  $n_s\approx 1-{2\over N}$, $r\approx   {12\alpha\over N^2}$ \cite{Kallosh:2013hoa,Ferrara:2013rsa,Kallosh:2013yoa,Cecotti:2014ipa,Galante:2014ifa,Kallosh:2015zsa,Ferrara:2016fwe,Kallosh:2017ced,Kallosh:2017wnt}.  These predictions are consistent with the latest observational data for $\alpha < O(10)$.

In the single-field realizations, the universality of these predictions can be ultimately traced back to the radial stretching introduced by the geometry (\ref{geometry}) as we approach the boundary $\rho \sim 1$.  On the other hand it is clear that, in the two-field embedding in terms of $Z$, the stretching also affects the ``angular" $\theta$-direction and this begs the question whether perhaps there is a regime where the predictions for the inflationary observables are also fairly insensitive to the details of the angular dependence of the potential.  In this paper we answer this question in the affirmative for sufficiently small $\alpha\lesssim O(1)$.

A particularly interesting case is $\alpha=1/3$, where a class of supergravity embeddings are known to possess an additional symmetry, which makes both $\rho$ and $\theta$ light \cite{Kallosh:2015zsa}. This means we cannot integrate out the angular field and we have to take into account the full multi-field dynamics. We will show that, in contrast with the naive expectation, the cosmological predictions of the simplest class of such models are very stable not only with respect to modifications of the potential of the field $\rho$, but also with respect to strong modifications of the potential of the field $\theta$. Importantly, we have to account for the full multi-field dynamics \cite{Gordon:2000hv, GrootNibbelink:2000vx, GrootNibbelink:2001qt, Bartolo:2001rt, Lalak:2007vi, Achucarro:2010jv, Achucarro:2010da, Peterson:2010np} in order to obtain the right results\footnote{See \cite{Welling:2015bra} for a recent review and references there.}.  The predictions coincide with the predictions of the single-field $\alpha$-attractors for $\alpha = 1/3$: $n_s\approx 1-{2\over N}$, $r\approx   {4\over N^2}$.
It was emphasized in \cite{Kallosh:2015zsa} that for $3\alpha=1$, the geometric kinetic term
 \be
{dZ d\bar Z\over (1-Z\bar Z)^2}
 \ee
 has a fundamental origin from maximal $\cN=4$ superconformal symmetry and from  maximal $\cN=8$ supergravity. Also  the single unit size disk, $3\alpha=1$,  leads to the lowest B-mode target which can be associated with the maximal supersymmetry models, M-theory, string theory and N=8 supergravity,  see \cite{Ferrara:2016fwe,Kallosh:2017ced} and \cite{Kallosh:2017wnt}.

More generally, we will also show that, for sufficiently small values of $\alpha < O(1)$,
the class of potentials exhibiting universal behaviour becomes very broad, and in particular it includes potentials with ${1 \over \rho} V_\theta \sim V_\rho \sim V$.

Our results lend support to the tantalizing idea, recently explored in some detail in \cite{Achucarro:2016fby} and building on earlier works in \cite{Kobayashi:2010fm, Turzynski:2014tza, Cremonini:2010sv, Cremonini:2010ua, vandeBruck:2014ata}, that multi-field inflation on a hyperbolic manifold may be compatible with current observational constraints {\it without the need to stabilize all other fields besides the inflaton.}   Since axion-dilaton moduli systems with the geometry \eqref{geometry} are ubiquitous in string compactifications, this observation could have important implications for inflationary model building.

Although at first sight the universality found here resembles a similar result obtained in the theory of multi-field conformal attractors \cite{Kallosh:2013daa} for $\alpha = 1$, the reason for our new result is entirely different. 
In the model studied in \cite{Kallosh:2013daa}, the light field $\theta$ evolved faster than the inflaton field, so it rapidly rolled down to the minimum of the potential with respect to the field $\theta$, and the subsequent evolution became the single-field evolution driven by the inflaton field. The observable e-folds are in the latter, single-field regime.
On the other hand, in the class of models to be discussed in our paper,
 the angular velocity $\dot \theta$  is exponentially suppressed, due to the hyperbolic geometry, and inflation proceeds (almost) in the radial direction. The angular field will \textit{not} roll down to its minimum, but instead it is "\textit{rolling on the ridge}". This is illustrated in Figures \ref{fig:stream} and \ref{rolling}.
 Nevertheless, the trajectory is curved and the inflationary dynamics is truly multi-field.

Multi-field models of slow-roll inflation based on axion-dilaton systems have been studied for some time \cite{Starobinsky:2001xq, DiMarco:2002eb}.
However, it is only fairly recently that the very important role played by the hyperbolic geometry for multi-field inflation is being recognized (see, e.g. \cite{Kallosh:2015zsa, Turzynski:2014tza, Renaux-Petel:2015mga, Ellis:2014opa, Achucarro:2016fby, Brown:2017osf, Mizuno:2017idt, Achucarro:2018def}). Unlike in previous works, here we choose to be agnostic about the potential, and derive the conditions that will guarantee universality of the inflationary predictions for the two-field system.

The paper is organized as follows.
In Section \ref{sugra implementation} we present a new supergravity embedding of the $\alpha = 1/3$ two-field model  with a light, non-stabilized, angular field, as an anti-D3 brane induced geometric inflationary model. We study its inflationary dynamics, and elaborate on the "rolling on the ridge" behaviour in Section \ref{sec:rolling}.
Next, we work out the universal predictions for primordial perturbations in Section \ref{sec:universal}, and leave the details of the full multi-field analysis for Appendix \ref{app:fullanalysispt}.
We extend this result to general values of $\alpha$ and work out the constraints on the potential to ensure the universality of the predictions in Section \ref{sec:general} and Appendix \ref{app:conditionspotential}.
Section \ref{sec:conclusions} is for summary and conclusions.

\section{\boldmath{$\alpha$-attractors and their supergravity implementations}}
\label{sugra implementation}

\begin{figure}[t]
\begin{center}
\includegraphics[scale=0.5]{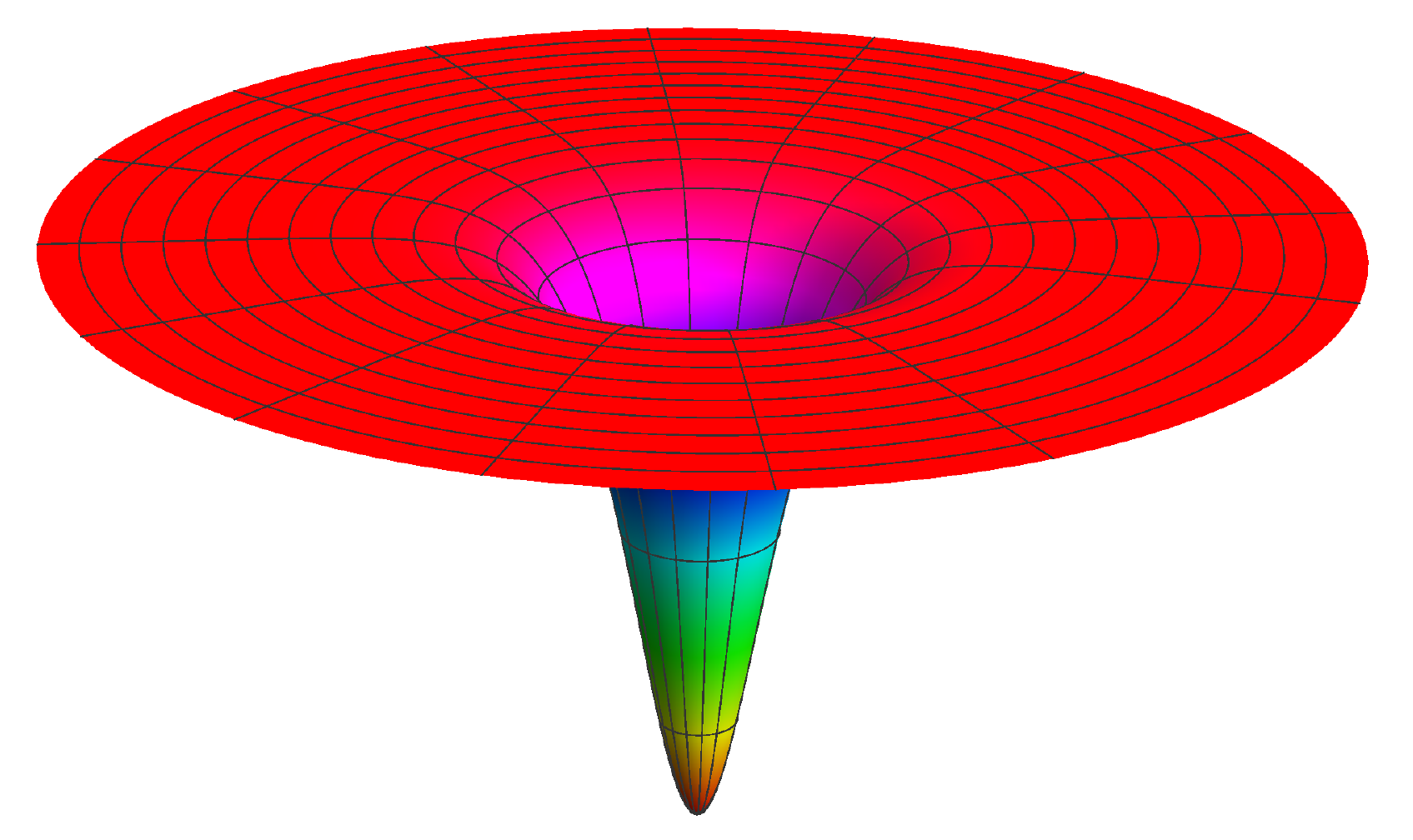}
\end{center}
\caption{\footnotesize The $\theta$-independent $3\alpha=1$ T-model  potential $V(\vp) = m^{2 }\tanh^2{\frac{\vp}{\sqrt 2}}\ $ . }
\label{Tmodel}
\end{figure}

There are several different formulations of $\alpha$-attractors in supergravity. One of the first formulations  {\cite{Kallosh:2013yoa}} was
based on the theory of a chiral superfield $Z$ with the \K\ potential corresponding to the Poincar\'e disk of  size $3\alpha$,
\be\label{k1}
K=-3\alpha \ln (1-Z\bar Z -S\bar S) \ ,
\ee
and superpotential
\be\label{w1}
W= S\, f(Z) (1-Z^{2})^{3\alpha-1\over 2} \ ,
\ee
where $f(Z)$ is a real holomorphic function. It is possible to make the field $S$ vanish during inflation, either by stabilizing it, or by making it nilpotent  \cite{Ferrara:2014kva}. Either way, the kinetic term for $Z$ is
 \be
 3\alpha {dZ d\bar Z\over (1-Z\bar Z)^2} .
 \ee
The field $Z$ can be represented as $e^{i\, \theta} \, \tanh{\vp\over \sqrt{6\alpha}}$, where $\vp$ is a canonically normalized inflaton field. In the simplest models of this class, the mass of the field $\theta$ in the vicinity of $\theta = 0$  during inflation 
is given by
\be
m^2_\theta= 2V\ \Big(1- {1\over 3\alpha}\Big),
\label{mass1}\ee
up to small corrections proportional to  slow roll parameters. In particular, for the simplest models with $\alpha > 1/3$ one finds $m^2_\theta>0$, which means that the field $\theta$ is stabilized at $\theta = 0$. Meanwhile for $\alpha > 2/5$ one has $m^2_\theta= V/3 \geq  H^{2}$ where $H$ is the Hubble constant. This means that the field $\theta$ for $\alpha \geq 2/5$
is strongly stabilized, and the only dynamical field during inflation 
is the inflaton field $\vp$ with the potential
\be\label{pot1}
V= \big |f(\tanh{\vp\over \sqrt{6\alpha}})\big |^{2}.
\ee
Meanwhile for $3\alpha \approx 1$ one finds that during inflation  $|m^2_{\theta}| \ll H^{2}$.  As an example,  the potential $V$ for $f(Z)  = m Z$ does not depend on $\theta$ at all:
\be
V= m^{2} \tanh^{2}{\vp\over \sqrt{6\alpha}} \ ,
\ee
see Figure~\ref{Tmodel}.

Later on, it was found  \cite{Carrasco:2015uma} that one can strongly stabilize the field $\theta$ for all $\alpha$ and reduce investigation of the cosmological evolution to the study of the single inflaton field in the models  with a somewhat different \K\ potential,
\be
K=-{3\alpha }  \ln {1 -Z\bar Z \over |1-Z^2|}  + S\bar S \ ,
\label{Cec3}\ee
and  superpotential
\be
W= S\, f(Z)   \ ,
\ee
which yields the same inflaton potential \rf{pot1} for $\theta = 0$.

This considerably simplifies investigation of inflationary models. An advantage of this \K\ potential is its manifest shift symmetry: it vanishes along the direction $Z = \bar Z$, corresponding to $\theta = 0$.

The next step was  the construction of the anti-D3 brane induced geometric inflationary models with arbitrary $\alpha$ with a stabilized field $\theta$  \cite{Kallosh:2017wnt} (see also \cite{McDonough:2016der}). The \K\, function is
\be
\cG=\ln |W_0|^2 -{3\alpha }  \ln {1 -Z\bar Z \over |1-Z^2|} +S +\bar S + G_{S\bar S}(Z, \bar Z) S\bar S \ ,
\ee
where  the field $S$ is nilpotent, with the metric
\be
G_{S\bar S}(Z, \bar Z) = {|W_0|^2 \over {\bf V} (Z, \bar Z)+ 3  |W_0|^2 } \ .
\label{G}\ee
The bosonic part of the supergravity action is
\be
g^{-1}{\cal L}= 3\alpha{dZ d\bar Z\over (1-Z\bar Z)^2} -  {\bf V} (Z, \bar Z) \ .
\label{susyAction}\ee
Note that the $Z$-part of the \K\, potential has the inflaton  shift symmetry at $Z=\bar Z$, as was shown in \cite{Carrasco:2015uma}.
The potential is
\be
{\bf V} (Z, \bar Z)= V(Z, \bar Z) +|F_S|^2-  3|W_0|^2 = V(Z, \bar Z) +\Lambda \ .
\label{potential}\ee
Here, as in all models in \cite{Kallosh:2017wnt}, $V(Z, \bar Z)$ is a function of $Z$ and $\bar Z$ which is regular at the boundary $Z\bar Z=1$ and which vanishes at the  minimum at $Z=0$, so that
\be
{\bf V} (Z, \bar Z)\Big |_{Z=0}= |F_S|^2-  3|W_0|^2  \equiv \Lambda \ .
\label{CC}\ee
The scale of 
supersymmetry breaking due to the  nilpotent field $S$ is
$
 e^{\cG}  \cG_S \cG^{S\bar S} \cG_{\bar S}\big |_{Z=0}=  |F_S|^2 \ ,
$
 and the gravitino mass is
$
 m^2_{3/2}\Big |_{Z=0}= |W_0|^2 \ .
$.
The angular field in these models is heavy, by construction, inflation takes place at $Z=\bar Z$.

This formulation is valid for any $\alpha$. However, subsequent investigations have revived interest in the specific models with $3\alpha = 1$  corresponding to the unit size disk \cite{Kallosh:2015zsa}, and in the possibility to describe models originating from  merger of several unit size disks, which may lead to $\alpha$-attractors with $3\alpha = 1, 2, 3,..., 7$ \cite{Ferrara:2016fwe,Kallosh:2017ced,Kallosh:2017wnt}.  It has been argued that these models provide some of the better motivated targets for the future B-mode searches.  Therefore it would be interesting to revisit all versions of these models, including the original versions with light, non-stabilized fields $\theta$ \cite{Kallosh:2015zsa}, since such models may exhibit a greater degree of symmetry, as shown in Figure~\ref{Tmodel}. It would be particularly interesting to find the corresponding generalization of the anti-D3 brane induced geometric inflationary models described above, applicable specifically to models with $3\alpha = 1$.

We find this new formulation by returning to the original \K\, frame with the axion shift symmetry, $K=-\ln (1-Z\bar Z)$, instead of
 $K= - \ln {1 -Z\bar Z \over |1-Z^2|}$. In this way the mass of the $\theta$-field will become light and we will have a two-field evolution on the disk of unit size $3\alpha=1$.
The \K\, function which provides the action
\be
g^{-1}{\cal L}= {dZ d\bar Z\over (1-Z\bar Z)^2} -  V(Z, \bar Z) -\Lambda
\label{susyAction1}\ee
will be taken in the following form:
\be
\cG=\ln |W_0|^2 -\ln(1-Z\bar Z) +S +\bar S + G_{S\bar S}(Z, \bar Z) S\bar S \ .
\ee
Here the metric of the nilpotent superfield is
\be
G_{S\bar S}(Z, \bar Z) = {|W_0|^2 \over (1-Z\bar Z)\Big ( |F_S|^2  +V(Z, \bar Z)\Big ) + 2|W_0|^2 Z\bar Z} \ .
\ee
It is different from the simpler version of $G_{S\bar S}$ in Equation \rf{G}, but the \K\,   potential $-\ln(1-Z\bar Z)$ as a function of $Z, \bar Z$  is simpler here. Moreover, the $Z$-part of the \K\, potential has an axion shift symmetry, it is $\theta$-independent.

One can show  that the expression for the scalar potential in this theory is given by
\be
{\bf V} (Z, \bar Z)= V(Z, \bar Z) +|F_S|^2-  3|W_0|^2 = V(Z, \bar Z) +\Lambda \ .
\label{potential2}\ee
This result is very similar to Equation \rf{potential}. However, \rf{potential} correctly represents the inflaton potential only along the inflaton direction $Z = \bar Z$. The potential for general values $Z \not = \bar Z$ must be calculated by the standard supergravity methods. This complication usually is not important for us since during inflation one can stabilize the fields along the inflaton direction $Z = \bar Z$.
Meanwhile in our new approach, equation \rf{potential2} gives the full expression for ${\bf V} (Z, \bar Z)$, which is valid for any    $Z$ and $\bar Z$ on the disk. This is a very special feature of the new formulation, which is valid for $3\alpha=1$.  

 During inflation, one can safely ignore the tiny cosmological constant $\Lambda \sim 10^{{-120}}$, so the potential \rf{potential2} is given by an arbitrary real function $V(Z, \bar Z)$. In the simplest cases, where $V$ is a  function of  $Z \bar Z$, it does not depend on the angular variable $\theta$, just as the potential in the theory \rf{k1} \rf{w1} for $3\alpha = 1$ shown in Figure \ref{Tmodel}. For more general potentials,  $V$ may depend on $\theta$, and the potentials can be quite steep with respect to $\rho$ and $\theta$.

 The key feature of this class of models, as well as of the models \rf{k1} \rf{w1} for $3\alpha = 1$, is that they describe hyperbolic moduli space corresponding to the  \K\ potential $K=-\ln (1-Z\bar Z)$, with the metric of the type encountered in the description of an open universe,  see Equation \rf{action1} below.   As we will see, the slow roll regime is possible for these two classes of theories even for very steep potentials, because of the hyperbolic geometry of the moduli space.

\section{Dynamics of multi-field $\alpha$-attractors}
\label{sec:rolling}

Now we come to study inflation with the above theoretical construction. Our starting point is
\be
g^{-1}{\cal L}= {dZ d\bar Z\over (1-Z\bar Z)^2} -  V(Z, \bar Z) ~.
\label{susyAction2}
\ee
The complex variable on the disk can be expressed as
\be
Z= \rho \, e^{i\theta} ~,
\ee
where $\rho$ is the radial field and $\theta$ is the angular field. In general, the potential $V(\rho,\theta)$  in these variables can be quite complicated  and steep. For simplicity, in the following we assume the potential vanishes at the origin $Z=0$ and is monotonic  along the radial direction of the unit disk\footnote{We leave other interesting cases with non-monotonic potential, such as the Mexican hat potential, for future work~\cite{Linde:2018hmx}.}, {\it i.e.} $V_\rho\geq0$.
One natural possibility is $V_\rho\sim V_\theta/\rho\sim V$, which at first glance cannot yield sufficient inflation. However, the hyperbolic geometry of the moduli space  makes slow roll inflation possible even if the potential is quite steep.

\begin{figure}
    \centering
        \includegraphics[width=0.65\textwidth]{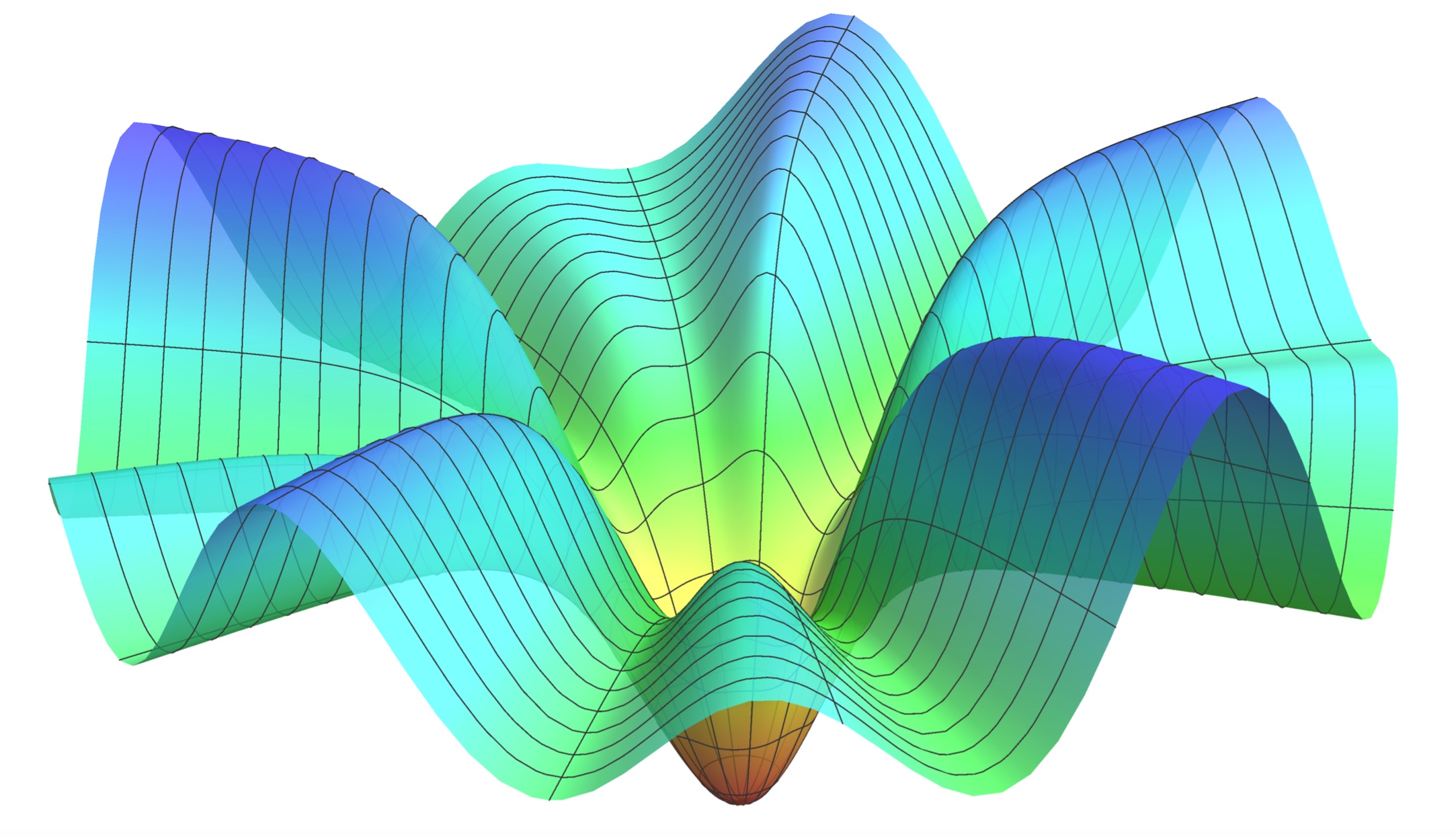}
        \caption{A stretched potential with angular dependence}\label{random}
\end{figure}

To see this, and to connect this to a more familiar canonical field $\vp$ in $3\alpha =1$ attractor models where the $\tanh$ argument is $\vp/\sqrt {6\alpha}$, we can use the following relation
\be
\rho=  \tanh {\vp\over \sqrt 2}  \ .
\ee
Therefore, our cosmological models with geometric kinetic terms are based on the following Lagrangian of the axion-dilaton system
\be
g^{-1}{\cal L} = {1\over 2}( \partial \vp)^2 + {1\over 4}\sinh ^2(\sqrt 2 \vp) (\partial\theta)^2 -V(\vp, \theta) \ ,
\label{action1}\ee
where  some  choice of the potentials $V(\vp, \theta)$ will be made depending on both moduli fields.  In terms of this new field $\vp$, the corresponding potential near the boundary $\rho = 1$ is exponentially stretched to form a plateau, where $\vp$ field becomes light and slow-roll inflation naturally occurs.
If we further assume the potential is a function of the radial field only, then we recover the T-model as shown in Figure~\ref{Tmodel}.
Generally speaking, the potential may also depend on $\theta$, and have ridges and valleys along the radial direction. One simple example is shown in Figure~\ref{random}. Although the $\theta$ field can 
appear heavy in the unit disk coordinates, after stretching in the radial direction, the effective mass in the angular direction is also exponentially suppressed for $\vp\gg1$.

For a cosmological spacetime, the background dynamics is described by equations of motion of two scalar fields
\be \label{eomvp}
\ddot\vp+3H\dot\vp+V_\vp-\frac{1}{2\sqrt2}\sinh\left(2\sqrt{2}\vp\right)\dot\theta^2=0~,
\ee
\be \label{eomth}
\ddot\theta+3H\dot\theta+\frac{V_\theta}{\frac{1}{2}\sinh ^2(\sqrt 2 \vp) }
+\frac{2\dot\theta\dot\vp}{\frac{1}{\sqrt2}\tanh(\sqrt2\vp)}=0~,
\ee
and the Friedmann equation
\be
3H^2 =   {1 \over 2} (\dot\vp^2 + {1 \over 2} \sinh^2{{\sqrt{2}\vp}} ~\dot \theta^2) + V(\vp, \theta) \ ,
\ee
where $H\equiv\dot{a}/a$ is the Hubble parameter.
In such a two-field system with potential as shown in Figure~\ref{random}, one may expect that the inflaton will first roll down from the ridge to the valley, and then slowly rolls down to the minimum along the valley. In the following we will demonstrate, 
due to the magic of hyperbolic geometry, the dynamics of moduli fields is totally different from this naive picture.

\subsection{Rolling on the ridge}
\label{ror}

In single-field $\alpha$-attractor models, inflation takes place near the edge of the Poincar\'e disk with $\rho\rightarrow1$ (or equivalently $\vp\gg1$).
Here we also focus on the large-$\vp$ regime where the potential in the radial direction is stretched to be very flat.
As a consequence, the radial derivative of the potential is exponentially suppressed
\be
V_\vp \simeq 2\sqrt2 V_\rho e^{-\sqrt2\vp}~.
\ee
After a quick relaxation, the fields can reach the slow-roll regime with the Hubble slow-roll parameters
\begin{equation} \label{epsilon}
 \epsilon \equiv -\frac{\dot{H}}{H^2}  = \frac{\dot\vp^2 + {1\over 2}\sinh ^2(\sqrt 2 \vp) \dot\theta^2}{2H^2} \ll 1~,~~~~\eta \equiv \frac{\dot\epsilon}{H\epsilon} \ll 1.
\end{equation}
Thus the kinetic energy of fields is much smaller than the potential, and the $\dot\theta\dot\vp$ term in \eqref{eomvp} is subdominant. Moreover, we assume that the field accelerations $\ddot\vp$ and $\ddot\theta$ can be neglected with respect to the potential gradient.
The equation of motion for $\theta$ is then simplified to 
\begin{equation}
 \frac{\dot{\theta}}{H} \simeq - {8}\frac{V_\theta}{V}{e^{-2\sqrt{2}\vp}}.
\end{equation}
This gives us the
velocity in the angular direction, which is highly suppressed in the large-$\vp$ regime.
Substituting the above result in the equation of motion for $\varphi$ \eqref{eomvp}, we can see that the centrifugal term 
proportional to $\dot\theta^2$ is also suppressed by $e^{-2\sqrt{2}\vp}$. Thus  for $\vp\gg1$ this term can be neglected compared to $V_\vp$. 
Therefore the equation of motion for $\vp$ is approximately
\be \label{srvp}
3H\dot\vp+V_\vp \simeq 0~,
\ee
which is the same as the single field case with slow-roll conditions.
Similarly we get the 
field velocity in the radial direction $\dot\vp\sim e^{-\sqrt2\vp}$, which is much larger than the angular velocity $\dot\theta$.
This is the main reason for the difference between the slow-roll regime in the present set of models, and in  the multi-field conformal attractors studied in \cite{Kallosh:2013daa}. In the conformal attractors, the field $\theta$ was rapidly rolling down, whereas here instead of rolling down to the valley first, the scalar fields are
\textit{rolling on the ridge} with {almost} constant $\theta$.

\begin{figure}[t]
  \centering
        \includegraphics[width=0.5\textwidth]{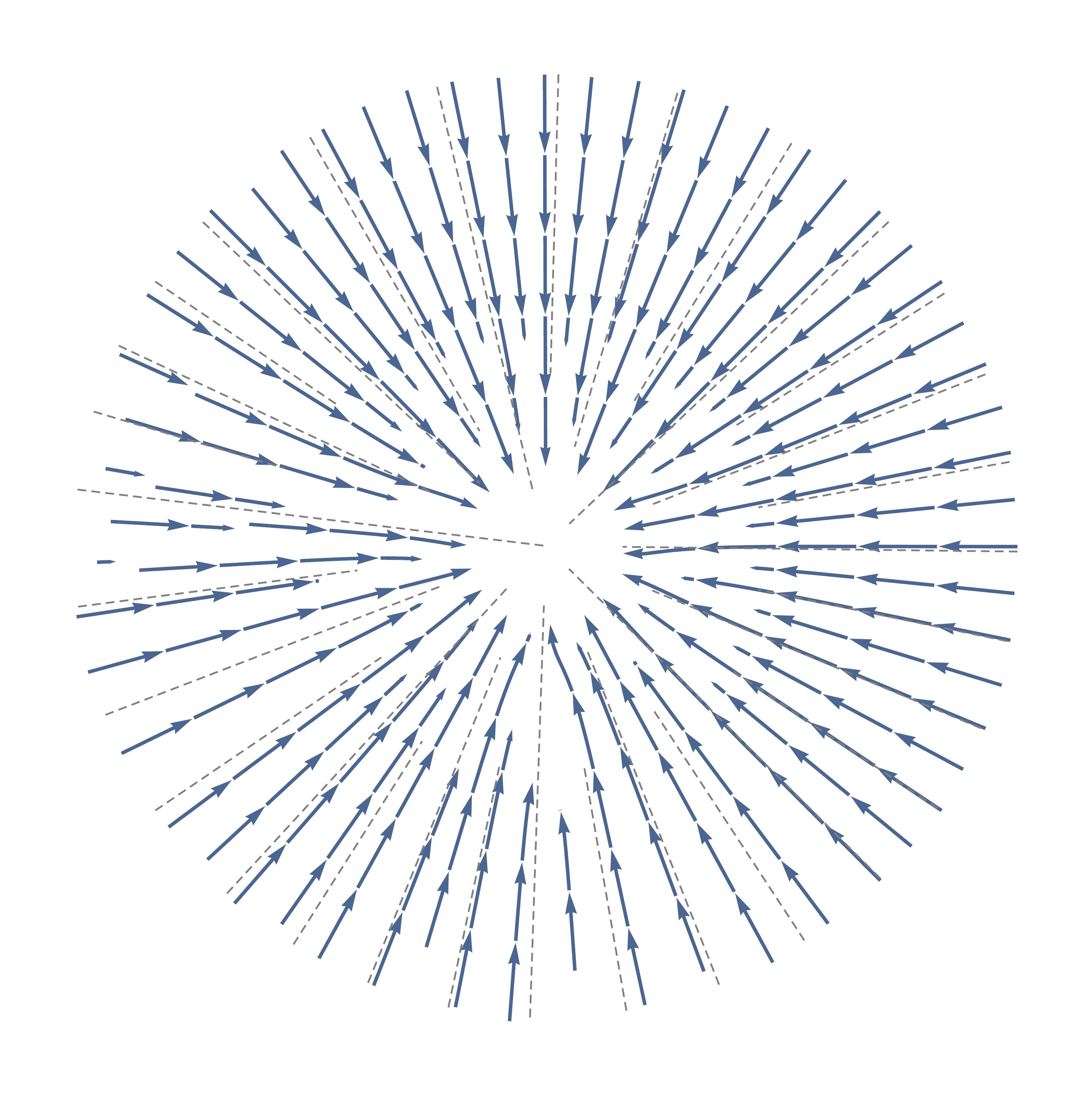}
        \caption{The stream of $\vp$ and $\theta$ fields on the potential with random angular dependence shown in Figure~\ref{random}. The dashed gray lines show the radial directions, while the blue arrows correspond to the field flow, starting at $\vp_i=10$.}
        \label{fig:stream}
\end{figure}

To see this counter-intuitive behaviour clearly, we can look at the flow $(\dot\vp,~\dot\theta)$ in the polar coordinate system.
The numerical result of the flow of the fields is shown in Figure~\ref{fig:stream} for the potential from Figure~\ref{random}.
As we see, although the potential {\em looks} chaotic in the angular direction, the fields always roll to the minimum along the ridge, no matter where they start.

However, it is 
crucial to emphasize that, although $\dot\theta$ is highly suppressed and $\theta$ is nearly constant, the angular motion is still quite important.
In the curved field manifold, since the angular distance is also stretched for large $\vp$, the 
proper velocity in the angular direction is given by ${1\over \sqrt2}\sinh (\sqrt 2 \vp)\dot\theta$.
We are encouraged to define a new parameter $\gamma$ as the ratio between 
the {\it physical} angular and radial velocity
\be \label{gamma}
\gamma\equiv\frac{\sinh (\sqrt 2 \vp)\, \dot\theta}{\sqrt2\, \dot\vp} \simeq \frac{V_\theta}{V_\rho} ,
\ee
where in the last step we have used large-$\vp$ and slow-roll approximations.
Since $\theta$ hardly evolves and $\rho\simeq1$ for $\vp\gg1$, $\gamma$ is nearly constant during most period of inflation.
This parameter captures the deviation from the single field scenario.
For instance, let us look at the potential slow-roll parameter in the radial direction
\be \label{eps}
\epsilon_\vp\equiv \frac12\(\frac{V_\vp}{V}\)^2 \simeq \frac{\dot\vp^2}{2H^2}~,
\ee
which is the same with the single field one.
Then in our model the full Hubble slow-roll parameter \eqref{epsilon} is approximately given by 
\be \label{epsgamma}
\epsilon=(1+\gamma^2)\epsilon_\vp~.
\ee
Thus a nonzero $\gamma$ demonstrates the contribution of the angular motion in the evolution of the two-field system.
Furthermore, depending on the form of the potential, $\gamma$ can be $\mathcal{O}(1)$ as we shall show in a toy model later. In such cases, the physical angular motion is comparable to the radial one, and the multi-field effects is particularly important\footnote{To see the importance of multi-field behaviour, another way is to look at the nonzero turning parameter, which we will discuss in Appendix \ref{app:fullanalysispt}.}.

In summary, for multi-field $\alpha$-attractors, there are two subtleties caused by the hyperbolic field space. First of all, the two-field evolution {\it looks} like the single field case without turning behaviour in the field space.
On the other hand, the straight trajectory is an illusion, and the {\it multi-field} effect can still be significant.
In Section \ref{sec:universal}, we will show how these surprising behaviours lead us to the universal predictions for primordial perturbations.

Concluding this subsection, we wish to further explain why "{\it rolling on the ridge}" is a quite general behaviour in multi-field $\alpha$-attractors.
Besides the aforementioned approximations, importantly we also
neglect the centrifugal term in \eqref{eomvp}.
Strictly speaking, this requires $V_\vp\gg\frac{1}{2\sqrt2}\sinh\left(2\sqrt{2}\vp\right)\dot\theta^2$,  which in the large-$\vp$ regime is equivalent to the following condition of the potential
\begin{equation} \label{condition}
\frac{V_\rho}{V}   \gg  \frac{4}{3}\left(\frac{V_\theta}{V}\right)^2  e^{-\sqrt{2}\vp} ~.
\end{equation}
Now we can see, near the boundary of the disk, unless the angular dependence of the potential is exponentially stronger than the radial one, the above condition always holds true and the system
evolves as we describe above. Finally let us stress that we have to ensure all our approximations are valid. We collect all conditions on the potential in Appendix \ref{app:conditionspotential}.
A natural choice of the potential with $V_\rho\sim V_\theta/\rho\sim V$ certainly 
satisfies these conditions.

\subsection{A toy model}
\label{sec:toy}

\begin{figure}
    \centering
        \includegraphics[width=0.6\textwidth]{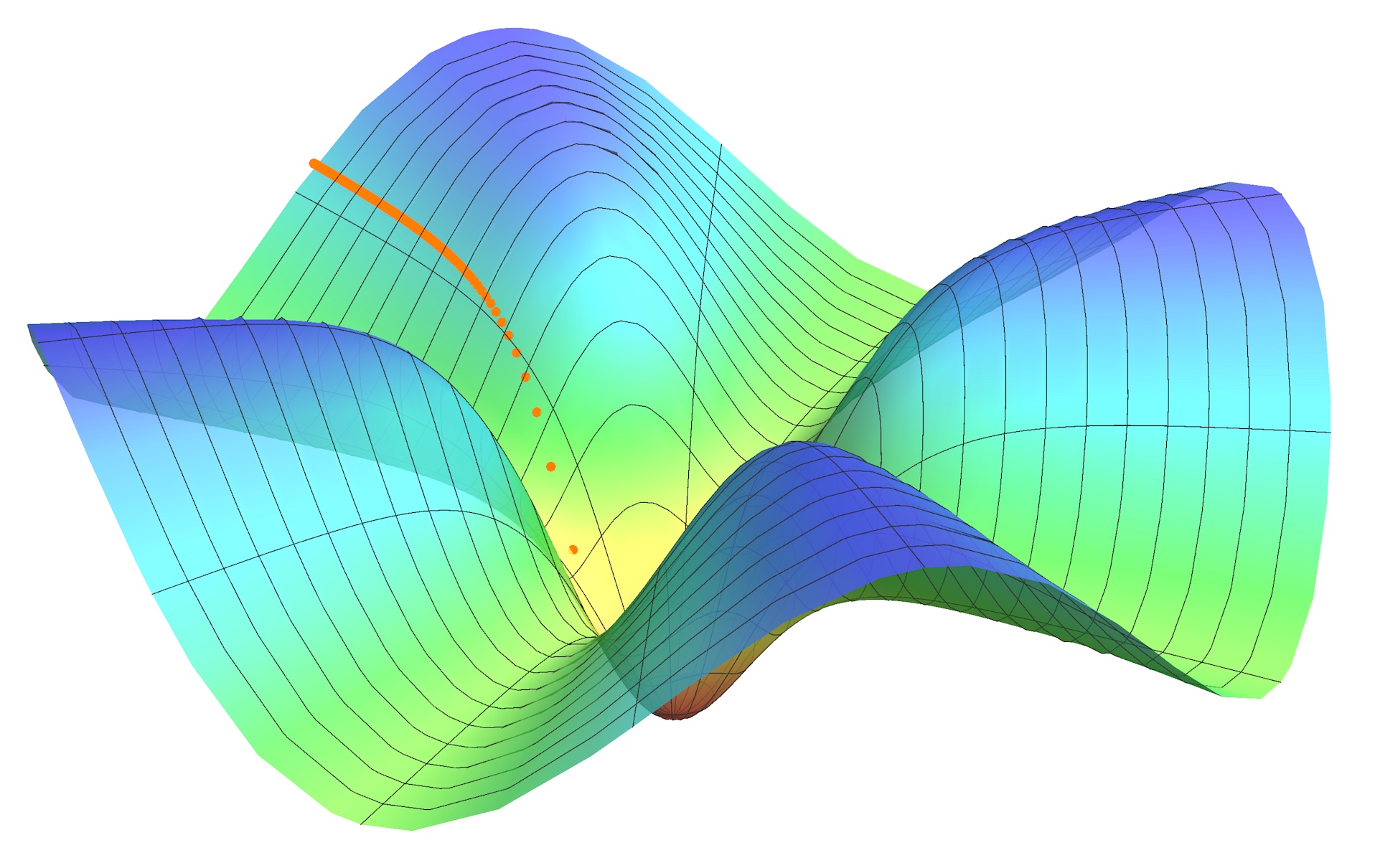}
        \caption{Rolling on the ridge: the form of the potential is given by the toy model \eqref{toymodelpotential} with $A = 0.2$, $n=4$ and initial angle $\theta_i =\pi/8$; the orange dots show a typical background trajectory, while the interval between the neighboring dots corresponds to one e-folding time.}\label{rolling}
\end{figure}

Before moving to the calculation of perturbations, let us work out a toy model to further confirm the above analysis.
Consider the following potential on the unit disk
\be
V(Z, \bar Z) =V_0\left[ Z\bar Z + A(Z^n+\bar{Z}^n)\right]~.
\ee
To ensure that  it is monotonic in the radial direction of the unit disk we need $A \leq \frac{1}{n}$. Then the condition \eqref{condition} is certainly satisfied.
In terms of $\vp$ and $\theta$, the potential is given by
\be \label{toymodelpotential}
V(\vp, \theta)=V_0 \tanh ^2\left(\frac{\vp}{\sqrt{2}}\right)\left[1+2A\cos (n\theta)\tanh ^{n-2}\left(\frac{\vp}{\sqrt{2}}\right) \right]~.
\ee
For demonstration, in the following we take $n=4$, $A=0.2$ and the initial angle  $\theta_i=\pi/8$.
We solve the background evolution of this two field system numerically.
Figure~\ref{rolling} shows the field trajectory on the toy model potential. We can see that the inflaton is rolling on the ridge with nearly constant $\theta$.

\begin{figure}
\centering
		 \includegraphics[width=0.47\textwidth]{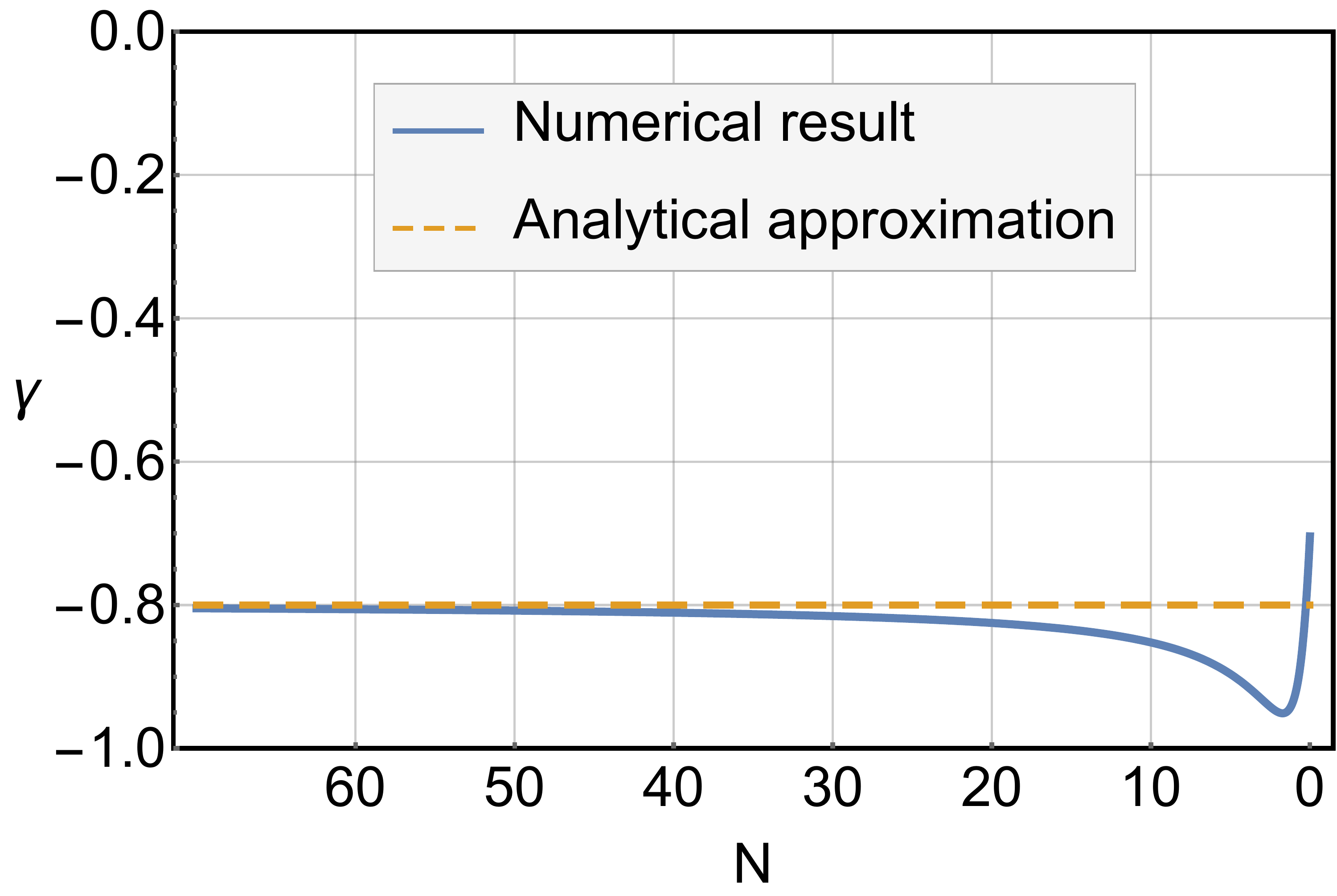}~~
               \includegraphics[width=0.475\textwidth]{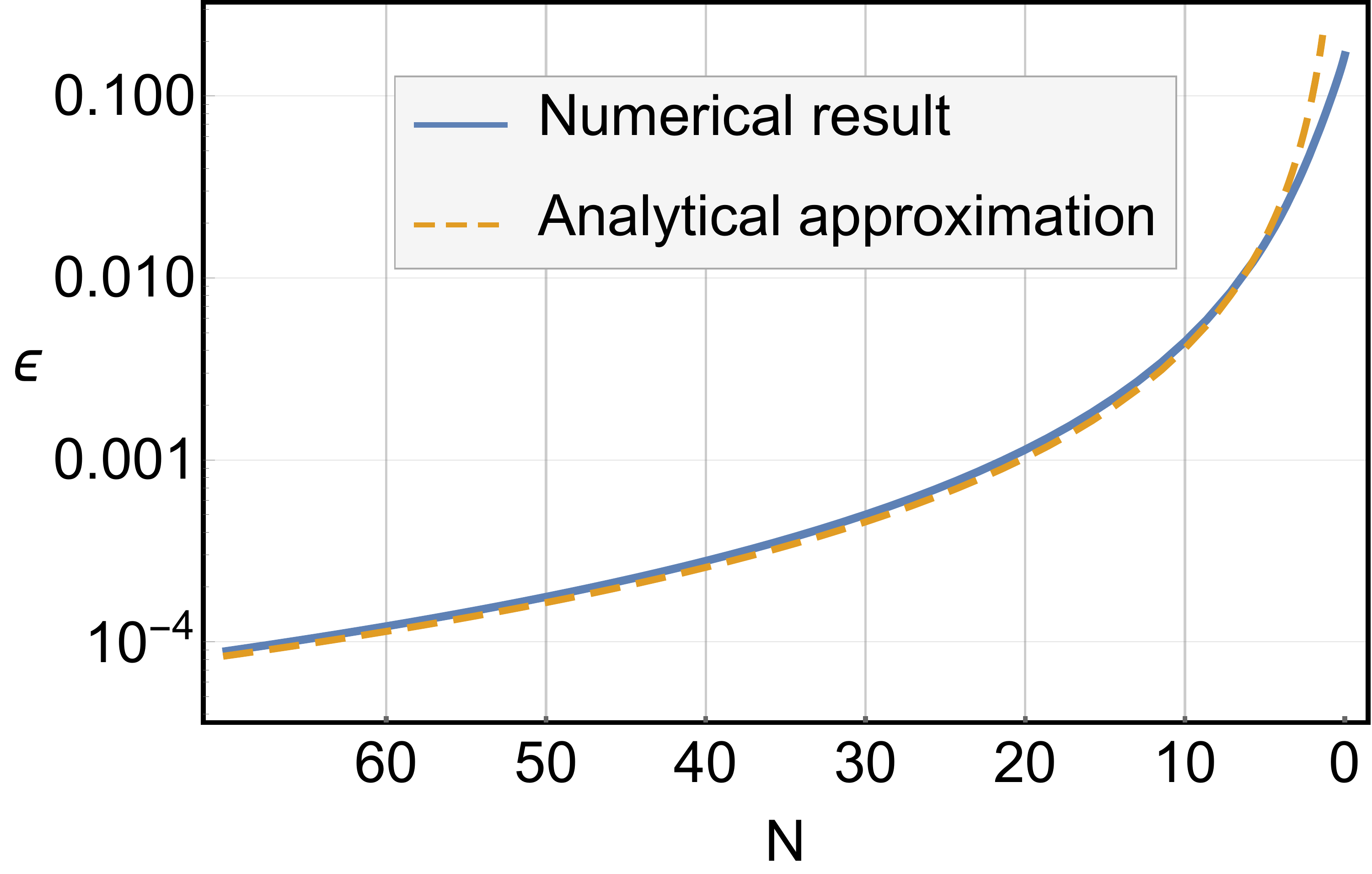}
        \caption{The evolution of $\gamma$ and $\epsilon$ in the toy model \eqref{toymodelpotential} with $A = 0.2$, $n=4$ and initial angle $\theta_i =\pi/8$.}
        \label{fig:gamma}
\end{figure}

Using the full numerical solution, we can check the validity of the large-$\vp$ and slow-roll approximations by looking at the  evolution of background parameters.
For example, within our analytical treatment, the $\gamma$ parameter is given by \eqref{gamma} as
\be
\gamma \simeq - \frac{nA\sin(n\theta)}{1+nA\cos(n\theta)}~.
\ee
It is nearly constant, since $\theta\simeq\theta_i$ during inflation. And the above choice of parameter values gives us $\gamma\simeq-0.8$, which agrees well with the numerical result as shown in Figure~\ref{fig:gamma}.
Next, let us look at the slow-roll parameter $\epsilon$. Solving \eqref{srvp} gives us its behaviour in terms of e-folding number as
\be
\epsilon \simeq \frac{1+\gamma^2}{4N^2}~,
\ee
where  \eqref{epsgamma} is used.
As shown in Figure~\ref{fig:gamma}, this provides a good approximation until the last several e-foldings of inflation, where $\vp\gg1$ is not valid any more.
It is interesting to notice that, during inflation the scalar field mainly rolls in the large-$\vp$ region, outside of which inflation would end very quickly. Therefore, the approximation $\vp\gg1$ does give a good description for the background dynamics.  In the following section and in Appendix \ref{app:fullanalysispt}, we will come back to this toy model, and use it as an example to demonstrate other aspects of multi-field $\alpha$-attractors.

\section{Universal predictions of $\alpha$-attractors}
\label{sec:universal}

One of the most important properties of single field $\alpha$-attractor inflation is the universal prediction for observations. For $\alpha\lesssim \mathcal{O}(1)$ and a broad class of potentials, as long as $V(\rho)$ is non-singular and rising near the boundary of the Poincar\'e disk,
the resulting scalar tilt and tensor-to-scalar ratio converge to
\begin{equation} \label{universal}
n_s =1  -\frac{2}{N} \quad \text{and} \quad r = \frac{12\alpha}{N^2},
\end{equation}
where $N\sim 50-60$ is the number of e-folds for 
modes we observe in the CMB.

\begin{figure}
\centering
        \includegraphics[width=0.6\textwidth]{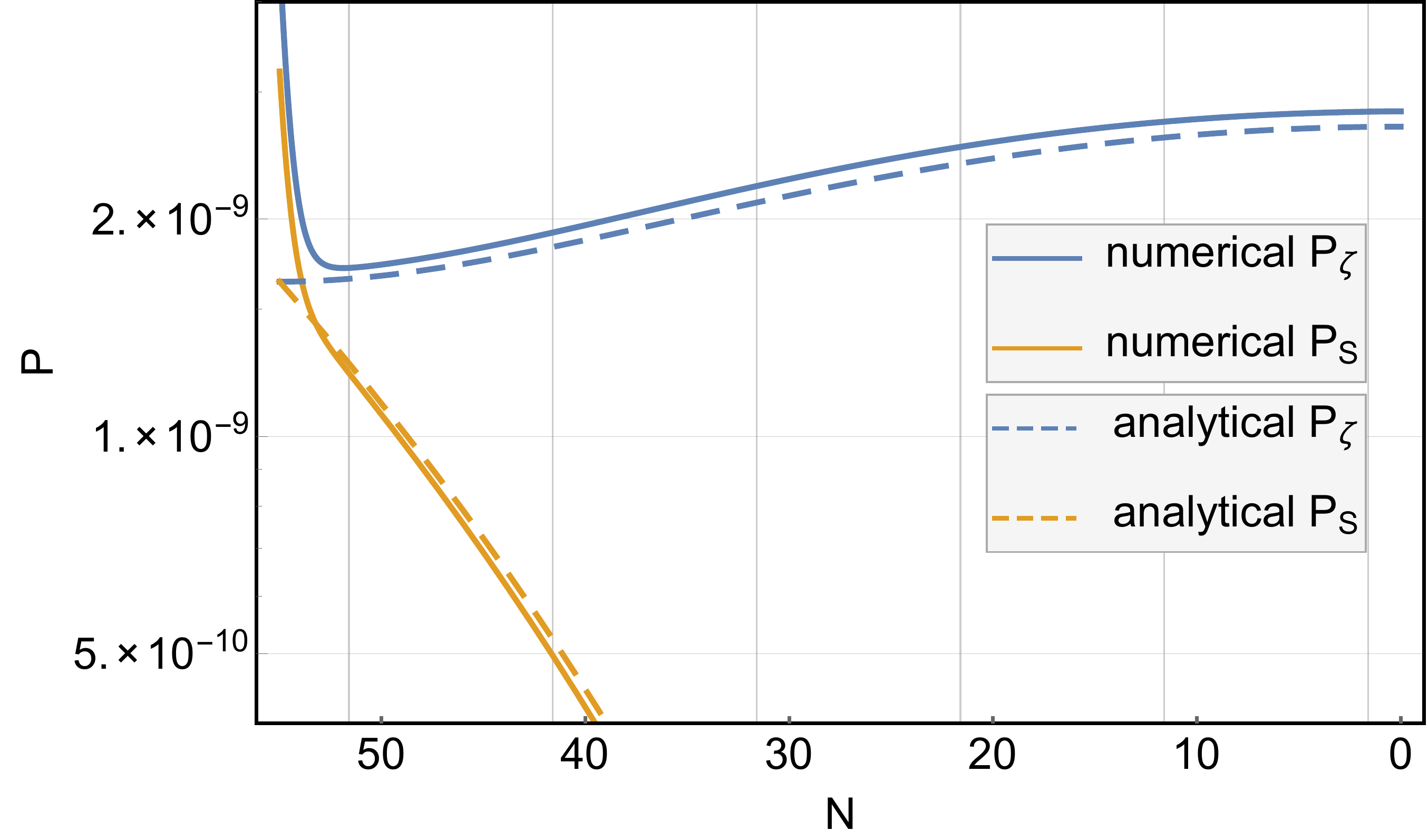}
        \caption{The evolution of curvature power spectrum $P_\zeta$ and isocurvature power spectrum $P_\mathcal{S}$ for perturbation
        modes which exit the horizon at $N=55$. We use the toy model \eqref{toymodelpotential} with $A = 0.2$, $n=4$ and initial angle $\theta_i =\pi/8$. The analytical solutions here are based on calculations in Appendix \ref{app:fullanalysispt}.}
        \label{fig:power}
\end{figure}

One interesting question is whether the universal predictions are still valid in the multi-field regime. In multi-field scenarios the curvature perturbation is sourced by the isocurvature modes on superhorizon scales, thus their evolution is typically non-trivial and yields totally different results for $n_s$ and $r$.
As we show above, 
the angular dependence in the $\alpha$-attractor potentials indeed leads to multi-field evolution.
For the toy model we studied, the behaviour of perturbations can be 
computed using the numerical code {\tt mTransport} \cite{Dias:2015rca}.
We focus on one single $k$ mode for curvature and isocurvature perturbations, and show their evolution in Figure~\ref{fig:power}. As expected, the curvature perturbation is enhanced during inflation, while the isocurvature modes decay.
Therefore, naively one expects there would be corrections to the single field $\alpha$-attractor predictions due to the multi-field effects.

In the following we will show that, surprisingly, the universal predictions are still valid in the multi-field regime.
We use the $\delta N$ formalism
to derive the inflationary predictions for the multi-field $\alpha$-attractor models studied in this paper. 
A full analysis of the perturbations is left for Appendix \ref{app:fullanalysispt}, where the evolution of
the coupled system of curvature and isocurvature modes is solved via the first principle calculation .

The $\delta N$ formalism \cite{Salopek:1990jq, Sasaki:1995aw, Starobinsky:1986fxa, Sasaki:1998ug, Lee:2005bb}
is an intuitive and simple approach to solve for the curvature perturbation in multi-field models.
At the end of inflation, regardless of the various field trajectories, the amplitude of curvature perturbations is only determined by the perturbation of the e-folding number $N$, which is caused by the initial field fluctuations.
Therefore, without studying details of the coupled system of curvature and isocurvature modes, as long as we know how the number of e-foldings $N$ depends on the initial value of the two fields, the curvature perturbation can be calculated.

Let us therefore consider how the initial  $\vp$ and $\theta$ determine $N$.
In this paper, we define the e-folding number as the one counted
backwards from the end of inflation, thus $dN=-Hdt$.
In terms of $N$, the slow-roll equation \eqref{srvp} becomes
\begin{equation}
 \frac{d\vp}{dN}  \simeq 2\sqrt{2} e^{-\sqrt{2}\vp}\frac{V_\rho}{V} ~.
\end{equation}
Since in the large $\vp$ regime $\rho\rightarrow1$ and $V_\rho/V$ is nearly constant for a given trajectory,
the equation above yields the e-foldings from the end of inflation as
\be \label{Nphi}
N=\frac{1}{B}e^{\sqrt{2}\vp} +C(\theta)~,
\ee
where $B\equiv4V_\rho/V$ and $C(\theta)$ is an $\mathcal{O}(1)$ integration constant which can be fixed by setting $N=0$ at the end of inflation.
Thus, both two fields affect the duration of inflation as 
expected in multi-field models.
By this expression, we can use the $\delta N$ formalism to find curvature perturbation at the end of inflation
\be \label{deltaN}
\zeta=\delta   N=\frac{\partial  N}{\partial \vp}\delta\vp + \frac{\partial  N}{\partial \theta}\delta\theta
= \frac{\sqrt2e^{\sqrt2\vp}}{B}\delta\vp
+\left(C_\theta-\frac{B_\theta}{B^2}e^{\sqrt2\vp}\right)\delta\theta~.
\ee
As we see here, $\frac{\partial  N}{\partial \vp}$ and $\frac{\partial  N}{\partial \theta}$ can be comparable to each other.
However, one should keep in mind that $\theta$ field is non-canonical, thus to estimate the field fluctuation amplitudes at horizon-exit, one should consider the canonically normalized ones: $\delta\vp$ and $\frac{1}{\sqrt2}\sinh(\sqrt2\vp)\delta\theta$. Approximately in the large-$\vp$ region we have the following relation
\be \label{suppress}
\delta\vp \simeq
\frac{e^{\sqrt2\vp}}{2\sqrt2}\delta\theta \simeq \frac{H}{2\pi}~.
\ee
From here, we find that the field fluctuation $\delta \theta$ 
is exponentially suppressed, compared to the one from $\delta\vp$. 
So we only need to take into account the first term in equation \eqref{deltaN}.
In addition, equation \eqref{eps} yields $\epsilon_\vp=B^2e^{-2\sqrt2\vp}/4$, which further simplifies the $\delta N$ formula to $\zeta\simeq \delta\vp/\sqrt{2\epsilon_\vp}$. In the end, the power spectrum of curvature perturbation can be expressed as
\be
P_\zeta\equiv \frac{k^3}{2\pi^2}{|\zeta_k|}^2 \simeq  \frac{H^2}{8\pi^2\epsilon_\vp}~.
\ee
It is interesting to note that this result does not depend on any parameters related to the multi-field effects (such as $\gamma$).
Using \eqref{eps} and \eqref{Nphi}, we also get $\epsilon_\vp\simeq 1/(4N^2)$, which has the same behaviour with the single-field potential slow-roll parameter $\epsilon_V$. Thus the power spectrum above is coincident with the single-field one.
Then the predictions of scalar tilt and tensor-to-scalar ratio follow directly
\begin{equation}
n_s-1 \simeq -\frac{2}{N} \quad \text{and} \quad r \simeq \frac{4}{N^2}~.
 \label{nsr}
\end{equation}
These results are further confirmed by solving the full evolution of perturbations as shown in Appendix \ref{app:fullanalysispt}.

The $\delta N$ calculation above also demonstrates the counter-intuitive properties of multi-field $\alpha$-attractors.
As we show in Section \ref{sec:rolling},
the stretching effects of hyperbolic geometry not only flattens the potential in the radial direction,
but also suppresses the angular velocity $\dot{\theta}$.
At the level of perturbations, the similar effect occurs to the field fluctuations in the angular direction.
While the canonically normalized angular field fluctuation
has the same amplitude with $\delta\vp$, the original
field perturbation $\delta\theta$ is exponentially suppressed.
Therefore, only the radial field fluctuation $\delta\vp$ contributes to the final result.

Furthermore, the above results do not depend on the initial values of $\theta$, which correspond to
different field trajectories as shown in Figure~\ref{fig:stream}.
Certainly  their respective e-folding number $N$ and $\epsilon_\vp$ can be different from each other.
However, the $N$-dependence of $\epsilon_\vp$ is the same for all the "rolling on the ridge" trajectories.
Thus regardless of various initial values of $\theta$, the multi-field $\alpha$-attractors yield the same universal predictions for $n_s$ and $r$.

Typically, another prediction in multi-field inflation is large local non-Gaussianity, which is disfavoured by the latest data \cite{Ade:2015ava}.
Therefore it is also worthwhile to estimate the size of the bispectrum in our model. Here we expand the $\delta N$ formula to the second order in field fluctuations
\be \label{dNnl}
\zeta=\delta   N=\frac{\partial  N}{\partial \vp}\delta\vp + \frac{\partial  N}{\partial \theta}\delta\theta + \frac{1}{2}\frac{\partial^2  N}{\partial \vp^2}\delta\vp^2 + \frac{1}{2}\frac{\partial ^2 N}{\partial \theta^2}\delta\theta^2 + \frac{\partial ^2 N}{\partial \theta \partial\vp}\delta\theta\delta\vp~.
\ee
In principle, there are two contributions here: one captured by the $\delta N$ expansion, and another one caused by field interactions of $\delta\vp$ and $\delta\theta$.
However, for the same reason shown in \eqref{suppress}, the $\delta\theta$-related terms in the expansion \eqref{dNnl} are highly suppressed. Moreover, since there is no large coupling between field fluctuations, we expect that the second contribution to the bispectrum will also be negligible.
As a result, the local non-Gaussianity is approximately given by $\delta\vp$ terms in \eqref{dNnl}
\be
f_{\rm NL}\simeq\frac{5}{6}\left.\frac{\partial^2  N}{\partial \vp^2}\right/\(\frac{\partial  N}{\partial \vp}\)^2\simeq\frac{5}{6N}~,
\ee
which is coincident with the single field consistency relation $f_{\rm NL}=-\frac{5}{12}(n_s-1)$~\cite{Maldacena:2002vr, Creminelli:2004yq}.
Again, we find the multi-field $\alpha$-attractor prediction returns to the single field one,
which further demonstrates the scope of universality.

\section {Universality conditions for more general $\alpha$}
\label{sec:general}

Our investigation was stimulated by the realization that $\alpha$-attractors have particularly interesting interpretation in supergravity models with $\alpha = 1/3$. A significant deviation  from $\alpha = 1/3$ typically either   makes the field $\theta$ tachyonic, or  strongly stabilizes it  at $\theta = 0$, which results in a single-field inflation driven by the field $\vp$, see e.g. \rf{mass1}.
One may wonder, however, what happens if we  consider a more general class of two-field $\alpha$-attractors, which may or may not have supergravity embedding, and concentrate on their general features related to the underlying hyperbolic geometry.

For general $\alpha$, the canonically normalized field in the radial direction is defined by
$\rho =   \tanh( {\vp/ \sqrt{6\alpha}}$),
which leads to the following kinetic term
\begin{equation}
 \frac{1}{2}\left(\partial \vp\right)^2 + \frac{3\alpha}{4} \sinh^2\left(\sqrt{\frac{2}{3\alpha}}\vp\right)\left(\partial \theta\right)^2.
\label{Iv1}
\end{equation}
The equations of motion \eqref{eomvp} and \eqref{eomth} also change accordingly, see \eqref{eomvpgeneralalpha} and \eqref{eomthgeneralalpha}.
Similarly as in Section \ref{rolling}, in the slow-roll and large-$\vp$ approximations 
these equations reduce to
\be
 \frac{\dot{\theta}}{H} \simeq -\frac{8}{3\alpha} \frac{V_\theta}{V} e^{-2\sqrt{2\over{3\alpha}} \vp}
~,~~~~~~3H\dot\vp\simeq- \frac{2\sqrt2}{\sqrt{3\alpha}} V_\rho e^{-\sqrt{2\over{3\alpha}} \vp}~.
\ee
As we see, the angular motion is also exponentially suppressed, compared to the radial one.
So the rolling on the ridge behaviour is not unique for $\alpha=1/3$, but quite general for any $\alpha\lesssim \mathcal{O}(1)$.
Similarly to \eqref{condition}, we get the following condition to ensure its validity
\begin{equation}
\frac{V_\rho}{V}   \gg  \frac{4}{9\alpha}\left(\frac{V_\theta}{V}\right)^2  e^{-\sqrt{2\over{3\alpha}} \vp} ~,
\label{conditiongeneralalpha}
\end{equation}
which can be  satisfied easily by many choices of potential, generalizing the bound \eqref{condition}   to other values of $\alpha$.
Therefore, the results follow directly just like we find in Section \ref{sec:rolling}.
For example, the ratio of proper velocities 
\be
\gamma = \frac{\sqrt{\frac{3\alpha}{2}}\sinh\left(\sqrt{\frac{2}{3\alpha}}\varphi\right)\dot\theta}{\dot\varphi}
\ee
is still nearly constant, while $\epsilon_\vp$ evolves as
\be
\epsilon_\vp\simeq \frac{3\alpha}{4N^2}~.
\ee
Repeating the same $\delta N$ calculation for perturbations, we get $N\simeq 3\alpha e^{\sqrt{2/3\alpha}\vp}/B$ and
\be
\zeta = \delta N\simeq \frac{1}{\sqrt{2\epsilon_\vp}} \delta\vp ,
\ee
which lead to the universal predictions \eqref{universal} for generic $\alpha$.
Therefore in a broader class of $\alpha$-attractors without supersymmetry, adding angular dependence to the potential will not modify the universal predictions either.  
Importantly, in order to validate the various assumptions we make to obtain the universal predictions, we need the potential to satisfy certain conditions.  The most non-trivial condition is already given in \eqref{conditiongeneralalpha}.  The additional constraints on the potential come from \textit{assuming} the slow-roll, `slow-turn' and large $\varphi$ approximation. We give more detail about these approximations and collect the constraints on the potential in Appendix \ref{app:conditionspotential}. Some of the conditions should also be satisfied for single field $\alpha$-attractors. 
 The smaller $\alpha$ becomes, the more pronounced the stretching of the hyperbolic field metric gets and it will be more likely to be within the large $\varphi$ regime $\vp \gtrsim \sqrt{\frac{3\alpha}{2}}$ and the slow-roll regime at the same time.  Finally, there are some additional constraints on the potential because of the multi-field nature of our class of models. In particular, if we want to have suppressed field accelerations, we need to satisfy the slow-roll and the slow-turn conditions given in \eqref{constraintetaphi} - \eqref{constraintturnphi}. A natural choice of the potential with $\frac{V_\theta}{\rho V} \sim \frac{V_\rho}{V} \sim \frac{V_{\theta\theta}}{\rho^2 V}\sim \frac{V_{\theta\rho}}{\rho V}\sim \frac{V_{\rho\rho}}{V} \sim 1$, evaluated at the boundary $\rho\sim1 $ amply satisfies all conditions for $\alpha \lesssim O(1)$. 

\section{Summary and Conclusions}
\label{sec:conclusions}

{In this paper we have studied the inflationary dynamics and predictions of a class of  $\alpha$-attractor models where both the radial and the angular component of the complex scalar field $Z= \rho\, e^{i\theta} $  are light during inflation.  We concentrated on the special case $\alpha=1/3$, where the model has a supergravity embedding with a high degree of symmetry from $\mathcal{N}=4$ superconformal or $\mathcal{N}=8$ supergravity. However, our results may have more general validity under the conditions specified in Appendix A.    Under the weak assumptions that the potential is monotonic in the radial coordinate, and the angular gradient is not exponentially larger than the radial gradient \eqref{condition},  we find exactly the same predictions as in the theory of the  single field $\alpha$-attractors:
\begin{equation}
n_s-1 \simeq -\frac{2}{N} \quad \text{and} \quad r \simeq \frac{12\alpha }{N^2} ~.
\end{equation}
 Universality of these predictions may make it difficult  to distinguish between different versions of $\alpha$-attractors by measuring $n_{s}$. However, from our perspective this universality is not a problem but an advantage of $\alpha$-attractors, resembling universality of several other general predictions of inflationary cosmology, such as the flatness, homogeneity and isotropy of the universe, and the flatness, adiabaticity and gaussianity of inflationary perturbations in single field inflationary models.}

The hyperbolic field metric plays a key role in finding these universal results. Let us summarize how we arrived at our new result and stress how the hyperbolic geometry dictates the analysis.
\begin{itemize}
\item
First, the hyperbolic geometry effectively stretches and flattens the potential in the radial direction to a shape independent of the original radial potential. Independent - as long as the potential obeys the condition \eqref{condition}. The amplitude of this shape, however, varies along the angular direction.

 \item  Next, the angular velocity $\dot \theta$  is exponentially suppressed, due to the hyperbolic geometry, and inflation proceeds (almost) in the radial direction. The inflaton
 is "\textit{rolling on the ridge}" in the $(\vp,~\theta)$ plane. This is illustrated in Figures \ref{fig:stream} and \ref{rolling}.

 \item The straight radial trajectory is an illusion, since the \textit{physical} velocity in the axion $\theta$ direction is typically of the same order as the radial velocity. The angle between the inflationary trajectory and the radial direction is nonzero and practically constant in this regime. Moreover, although the field is following the gradient flow, the trajectory is curved in the hyperbolic geometry. Therefore, the perturbations are coupled and the multi-field effects have to be taken into account.

 \item Then, we use the $\delta N$ formalism to compute the power spectrum of curvature perturbations (confirmed by a fully multi-field analysis in Appendix \ref{app:fullanalysispt}). The typical initial $\theta$ perturbations are very small and have a negligible effect on the number of efolds. However, the initial value of $\theta$ of a given trajectory determines how much a perturbation in the radial direction affects the number of efolds, since the trajectory is curved. At the same time the initial value of $\theta$ determines the renormalization of the slow-roll parameter $\epsilon$. {\it These two effects cancel exactly}, leaving us with the same predictions as the single field $\alpha$-attractors.
 Also the non-Gaussianity calculation recovers the single field result $f_{\rm NL} \simeq -\frac{5}{12}(n_s-1)$.

\item
 Finally, in Section \ref{sec:general} and Appendix \ref{app:conditionspotential}, we relax the condition   $\alpha = 1/3$  and simply assume the hyperbolic geometry (\ref{geometry}) with smooth potentials. We identify the conditions on the potential in order to exhibit the universal behaviour discussed in our paper, see \eqref{eqn:conditionsV}. For $\alpha \lesssim O(1)$ these conditions are amply satisfied by a broad class of potentials $V(\rho,\theta)$, including natural ones without a hierarchy of scales: 
$\frac{V_\theta}{\rho V} \sim \frac{V_\rho}{V} \sim \frac{V_{\theta\theta}}{\rho^2 V}\sim \frac{V_{\theta\rho}}{\rho V}\sim \frac{V_{\rho\rho}}{V} \sim 1$, evaluated at the boundary $\rho\sim1 $.

\end{itemize}

{In conclusion, the main result of our investigation is the stability of predictions of the cosmological $\alpha$-attractors with respect to significant modifications of the potential in terms of the original geometric variables $Z$. Whereas the stability with respect to the dependence of the potential on the radial component of the field $Z$ is well known \cite{Kallosh:2013yoa}, the stability with respect to the angular component of the field $Z$ is a novel result which we did not anticipate when we began this investigation.}

{Our results could have important implications for constructing UV completions of inflation. We have confirmed again that multi-field models of inflation can be perfectly compatible with the current data, in particular when the additional fields are very light. This lends support to the idea that it is not always necessary to stabilize all moduli fields in order to have a successful model of inflation.}

\

{
{\it Note added}: 
After submission, Ref.~\cite{Yamada:2018nsk} appeared, which provides a supergravity embedding for the more general $\alpha <1$ models discussed in Section \ref{sec:general}.
}

\

 \noindent{\bf {Acknowledgments:}}  We thank Sebasti\'an C\'espedes, Anne-Christine Davis, Oksana Iarygina, Gonzalo Palma, Diederik Roest, and Valeri Vardanyan for stimulating discussions and collaborations on related work.  The work  of  RK,  AL  is supported by SITP and by the US National Science Foundation grant PHY-1720397.   RK and  AL are grateful to the Lorentz Center in Leiden for the hospitality when this work was performed. DGW and YW are supported by a de Sitter Fellowship of the Netherlands Organization for Scientific Research (NWO). The work of AA is partially supported
by the Netherlands' Organization for Fundamental Research in Matter (FOM),  by the Basque Government (IT-979-16) and by the Spanish Ministry MINECO  (FPA2015-64041-C2-1P).

\begin{appendices}
\section{Constraints on the potential}
\label{app:conditionspotential}
In this Appendix we collect the conditions the potential has to obey in order to validate our approximations for any value of $\alpha$. Let us first recap some relevant definitions and equation for general $\alpha$. First of all, our three radial variables are given by $\varphi$ and
\begin{equation}
 R(\vp) \equiv \sqrt{\frac{3\alpha}{2}}\sinh\left(\sqrt{\frac{2}{3\alpha}}\varphi\right), \quad 
 \rho \equiv \tanh\left(\frac{\varphi}{\sqrt{6\alpha}}\right).
\end{equation}
We introduced the radial variable $R(\vp)$ because it appears naturally in the \textit{physical} angular velocity $R(\varphi) \dot\theta$.
The kinetic term can now be written in three equivalent ways
\begin{equation}
\frac{1}{2}\frac{(\partial \rho)^2 + \rho^2 (\partial \theta)^2}{(1-\rho^2)^2}  = \frac{1}{2}\left(\partial \vp\right)^2 + \frac{3\alpha}{4} \sinh^2\left(\sqrt{\frac{2}{3\alpha}}\vp\right)\left(\partial \theta\right)^2 =\frac{1}{2}(\partial \varphi)^2 + \frac{1}{2}R(\vp)^2 (\partial \theta)^2.
\end{equation}
The equations of motion are generalized to
\begin{equation}
 \ddot\vp+3H\dot\vp+V_\vp-\frac{1}{2}\sqrt{\frac{3\alpha}{2}}\sinh\left(2\sqrt{\frac{2}{3\alpha}}\vp\right)\dot\theta^2=0~,
 \label{eomvpgeneralalpha}
 \end{equation}
 \begin{equation}
  \ddot\theta+3H\dot\theta+\frac{V_\theta}{\frac{3\alpha}{2}\sinh^2\left(\sqrt{\frac{2}{3\alpha}} \vp\right) }
+\frac{2\dot\theta\dot\vp}{\sqrt{\frac{3\alpha}{2}}\tanh\left(\sqrt{\frac{2}{3\alpha}}\vp\right)}=0~.
\label{eomthgeneralalpha}
\end{equation}
Now we are ready to collect all constraints on the potential.
 In our derivation we \textit{assume} that we can neglect $\ddot{\varphi}$ and that we can take the large-$\varphi$ approximation. Moreover, it is important that we can neglect the centrifugal term proportional to $\dot\theta^2$ in Equation \eqref{eomvpgeneralalpha}. We use the gradient flow to estimate the size of $\dot{\theta}$, and this leads to the first constraint \eqref{constraintcentrifugal}.   For consistency, we have to ensure the validity of:
ncy, we have to ensure the validity of:
\begin{itemize}
 \item The slow-roll approximation, which gives rise to the next four constraints \eqref{constraintepsphi} -  \eqref{constraintetatheta}. This approximation ensures that the field velocities are small and that we can neglect their acceleration pointing along the corresponding field direction as well.
 
  \item  The slow-turn approximation, which allows us to neglect the field accelerations pointing along the \textit{other} field direction. If we can assume gradient flow for $\theta$ this leads to the condition \eqref{constraintturnphi}.
 
\item  The large-$\varphi$ approximation, which requires us not to go to the extreme limit of a very shallow radial potential. We want to inflate sufficiently far from the origin in order to obtain enough efolds of inflation, such that we can use the large-$\varphi$ approximation. In our analysis we work for simplicity with potentials  $\left(\frac{V_\rho}{V}\right)^2 \gtrsim  \frac{\alpha}{4}$ so this is automatically satisfied. 
\end{itemize}
\begin{subequations}
\begin{empheq}[box=\fbox]{align}
\frac{V_\rho}{V} &\gg \frac{4}{9\alpha} \left(\frac{V_\theta}{V}\right)^2 e^{-\sqrt{2/3\alpha}\vp}, \label{constraintcentrifugal}\\
 \epsilon_\vp &\equiv \frac{1}{2}\left(\frac{V_\vp}{V}\right)^2 = \frac{4}{3\alpha}\left(\frac{V_\rho}{V}\right)^2 e^{-2\sqrt{2/3\alpha}\vp}  \ll 1,\label{constraintepsphi} \\
 \epsilon_\theta &\equiv \frac{1}{2}\left(\frac{V_\theta}{R V}\right)^2 = \frac{4}{3\alpha}\left(\frac{V_\theta}{V}\right)^2 e^{-2\sqrt{2/3\alpha}\vp} \ll 1, \label{constraintepstheta}\\
  \eta_\varphi &\equiv \frac{1}{3}\frac{V_{\varphi\varphi}}{V} = \frac{8}{9\alpha}\frac{V_{\rho\rho}}{V} e^{-2\sqrt{2/3\alpha}\vp} \ll 1 \label{constraintetaphi}\\
  \eta_\theta &\equiv \frac{1}{3}\frac{V_{\theta\theta}}{R^2V} = \frac{8}{9\alpha}\frac{V_{\theta\theta}}{V} e^{-2\sqrt{2/3\alpha}\vp} \ll 1, \label{constraintetatheta}\\
 \omega_\phi &\equiv \frac{V_{\theta\varphi}}{3 R V}\frac{V_\theta}{R V_\varphi} = \frac{V_{\theta \rho}}{V} \frac{V_\theta}{V_\rho} \frac{8}{9\alpha} e^{-2\sqrt{2/3\alpha}\vp}\ll 1. \label{constraintturnphi}
\end{empheq}
\label{eqn:conditionsV}
\end{subequations}
Please note that all constraints have to be evaluated at $\rho \sim 1$, i.e. at $\vp \gg 6\alpha$.  Our conditions are satisfied for simplest  potentials, because in the large-$\varphi$ regime  all slow-roll and slow-turn parameters are exponentially suppressed. 
For instance, natural potentials which satisfy $\frac{V_\theta}{\rho V} \sim \frac{V_\rho}{V} \sim \frac{V_{\theta\theta}}{\rho^2 V}\sim \frac{V_{\theta\rho}}{\rho V}\sim \frac{V_{\rho\rho}}{V} \sim 1$ at the boundary $\rho \sim 1$, amply obey the conditions.

\section{Full analysis of perturbations}
\label{app:fullanalysispt}

In this Appendix, we give a detailed study of turning trajectories in multi-field $\alpha$-attractors and work out the full evolution of curvature and isocurvature  perturbations.

\subsection{Covariant formalism and large-$\vp$ approximations}

For a general multi-field system spanned by coordinate $\phi^a$ with field metric $G_{ab}$,
the equations of motion in the FRW background can be simply written as
\be
D_t\dot\phi^a +3H\dot\phi^a +V^a=0~,~~~~3H^2= \frac12\dot\Phi^2+V
\ee
where $D_t$ is the covariant derivative respect to cosmic time and $\dot\Phi^2\equiv G_{ab}\dot\phi^a\dot\phi^b$.
To describe the multi-field effects, it is convenient to define the tangent and orthogonal unit vectors along the trajectory as
\be
T^a\equiv \frac{\dot\phi^a}{\dot\Phi}~,~~~~ N_a\equiv \sqrt{\det G} \epsilon_{ab}T^b~,
\ee
where
 $\epsilon_{ab}$ is the Levi-Civita symbol with $\epsilon_{12}=1$.
The rate of turning for the background trajectory is defined as
\be
\Omega \equiv-N_aD_tT^a =\frac{V_N}{\dot\Phi}~,
\ee
where for the second equality we have used the background equations of motion and $V_N=N^a\nabla_aV$ is the gradient of the potential along the normal direction of the trajectory.
This quantity, which vanishes in single field models, is particularly important for the multi-field behaviour and evolution of perturbations.
A dimensionless turning parameter is introduced as
\be
\lambda\equiv -\frac{2\Omega}{H}~.
\ee
Another important parameter is the field mass along the orthogonal direction defined as
\be
V_{NN}\equiv N^aN^b\nabla_a\nabla_b V~.
\ee

\begin{figure}
    \centering
        \includegraphics[width=0.51\textwidth]{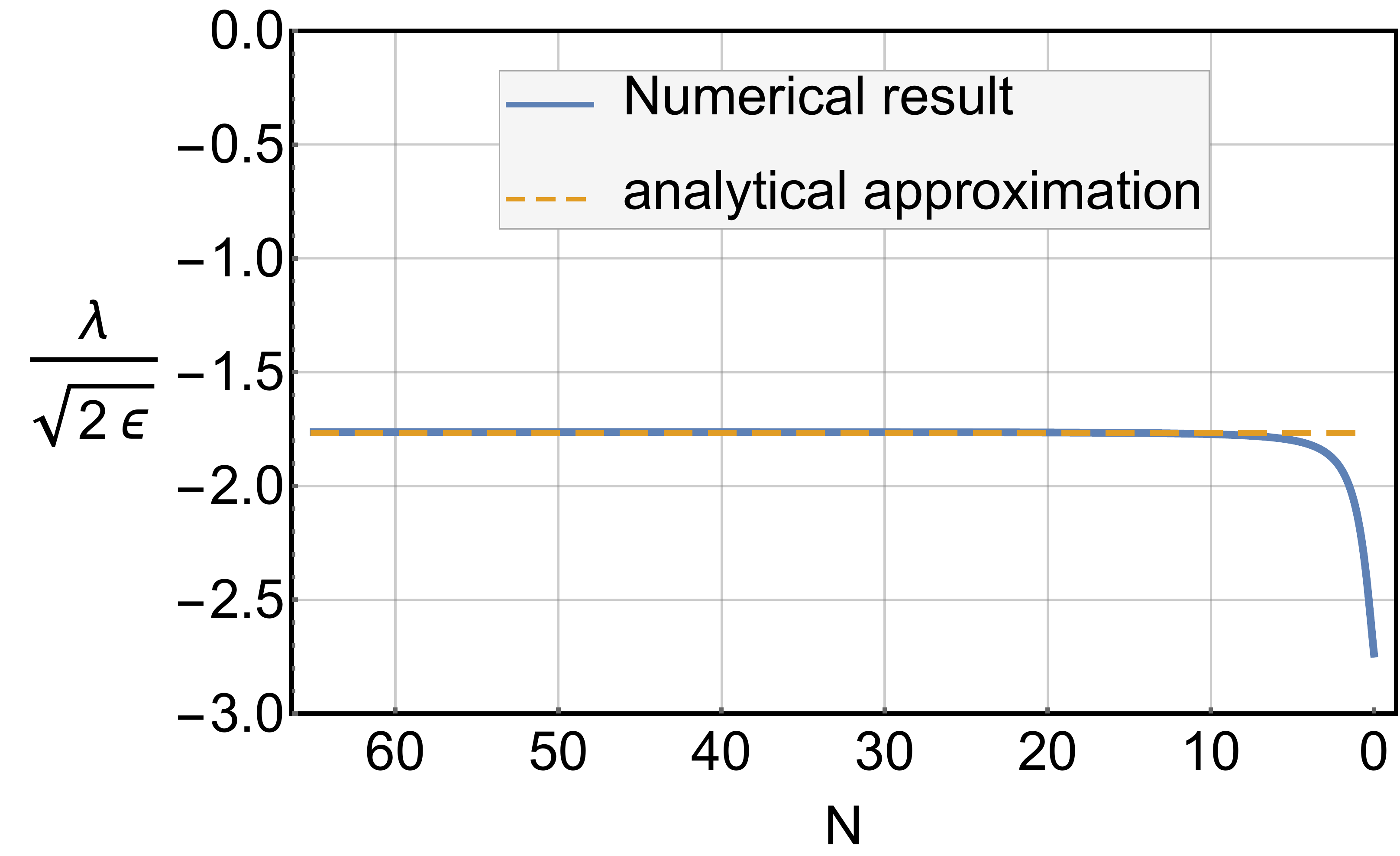}
        \includegraphics[width=0.47\textwidth]{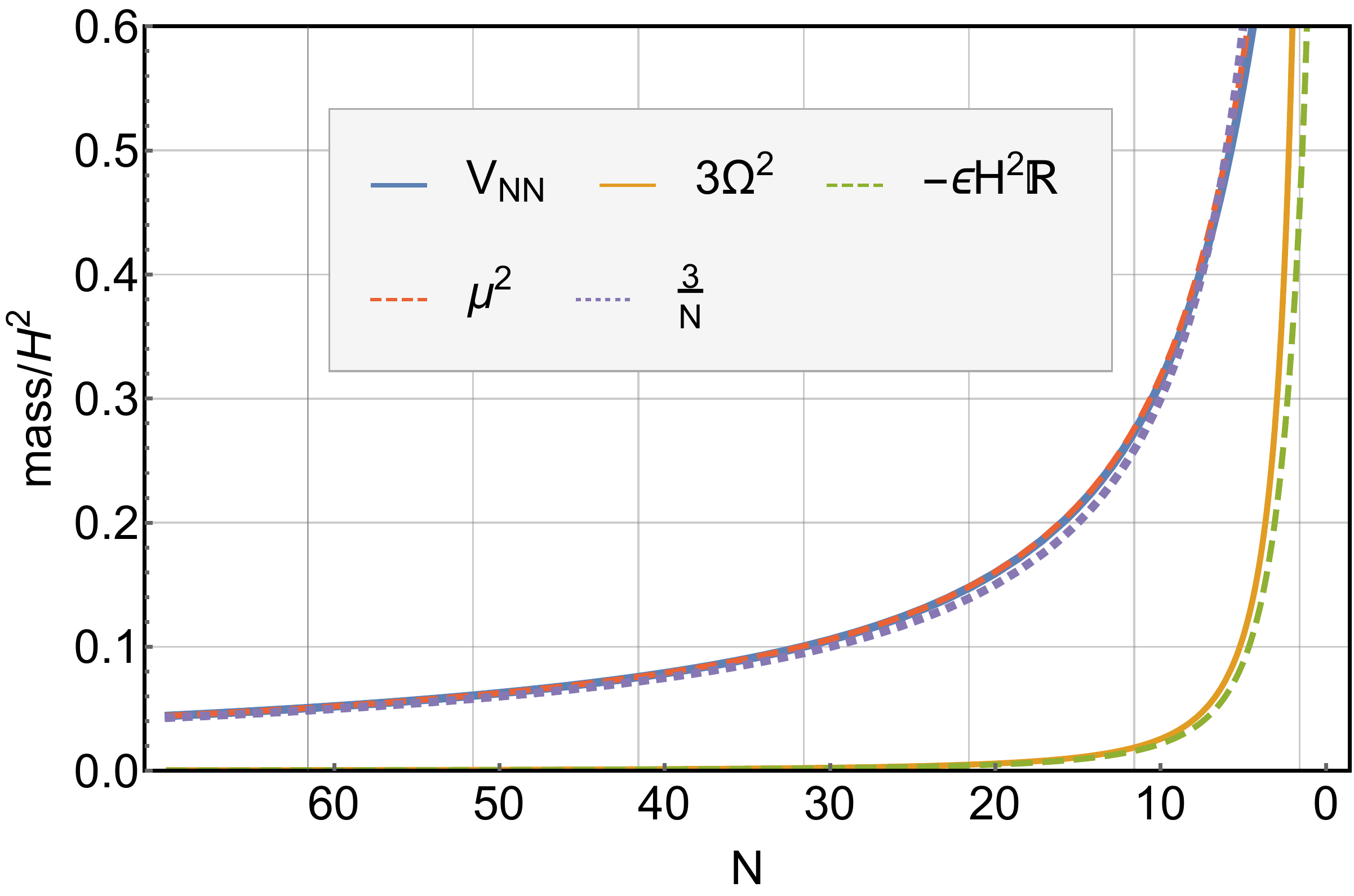}
    \caption{The evolution of the dimensionless turning parameter $\frac{\lambda}{\sqrt{2\epsilon}}$ and entropy masses. Here we use the toy model potential \eqref{toymodelpotential} with $n=4$, $A=0.2$ and $\theta_i=8/\pi$.}\label{fig:bglambdagamma}
\end{figure}

Now let us come back to our model with coordinates $\phi^a=(\vp, \theta)$ and hyperbolic field metric
\be
G_{ab}=
\begin{pmatrix}
1 ~ & ~ 0 \\
0 ~ & ~  {1\over 2}\sinh ^2(\sqrt 2 \vp)
\end{pmatrix}~.
\ee
The Ricci scalar of this manifold is a negative constant $\mathbb{R}=-2$.
By the definitions above, after some algebra, $\lambda$ and $V_{NN}$ here can be written into the following form
\be \label{lambda}
\lambda=\frac{1}{\epsilon H^3}{1\over \sqrt2}\sinh (\sqrt 2 \vp)\left[\ddot\vp\dot\theta-\ddot\theta\dot\vp-\frac{2\dot\theta\dot\vp^2}{\frac{1}{\sqrt2}\tanh(\sqrt2\vp)}-\frac{1}{2\sqrt2}\sinh\left(2\sqrt{2}\vp\right)\dot\theta^3\right]~,
\ee
\ba
V_{NN}&=& \frac{1}{\dot\Phi^2}\left(
\frac{V_{\theta\theta}\dot\vp^2 + \frac{\sqrt{2}}{4} \sinh(2\sqrt2\vp)V_\vp\dot\vp^2}{{1\over 2}\sinh^2 (\sqrt 2 \vp)}
+2\dot\theta\dot\vp\left[\frac{\sqrt{2}V_\theta}{\tanh(\sqrt2\vp)}-V_{\theta\vp}\right]\right.\nn\\
&&\left. ~~~~~~~~ +{1\over 2}\sinh^2 (\sqrt 2 \vp)V_{\vp\vp}\dot\theta^2 \right)~.
\label{eqn:VNN}
\ea
These expressions look very complicated, but in the large-$\vp$ regime they can be efficiently simplified.
First of all, since  $\gamma$ in \eqref{gamma} is nearly constant, we can use this parameter to replace $\dot\theta$ by $\dot\vp$ in these expressions, for example $\dot{\Phi}^2 = (1+\gamma^2)\dot{\vp}^2$.
Then we can use the relations of background quantities presented in Section \ref{rolling} to further simplify the  result. Finally the turning parameter $\lambda$ can be expressed as
\be
\lambda=\frac{-1}{\epsilon H^3} (1+\gamma^2) \frac{\sqrt{2}\gamma\dot\vp^3}{\tanh(\sqrt{2}\vp)}
\simeq \frac{2\sqrt{2}\gamma}{(1+\gamma^2)^{1/2}}\cdot \sqrt{2\epsilon}  ~,
\label{eqn:largephilambda}
\ee
where the large $\vp$ approximation is used in the last step. Therefore, at $\vp\gg1$, $\lambda/\sqrt{2\epsilon}$ is nearly constant.
Similarly, we can work out the approximated expression for $V_{NN}$. Here we use the toy model potential for demonstration, which yields
\begin{equation} \label{vnn2}
V_{NN} \approx V_0 B e^{-\sqrt{2}\vp}.
\end{equation}
Therefore, $V_{NN}$ is nearly zero at the beginning of inflation, but then grows up as $\vp$ rolls to the center. These analytical approximations are checked by using numerical solution of the toy model. In Figure~\ref{fig:bglambdagamma} we show the numerical results versus the analytical ones for $n=4$, $A=0.2$ and $\theta_i=8/\pi$. Indeed we see that
$\frac{\lambda}{\sqrt{2\epsilon}}$ remains constant until the very end of inflation, where the large-$\vp$ approximation breaks down.

\subsection{Primordial Perturbations}
With the analytical approximations developed above, now we can move to study the behaviour of  perturbations. In particular, we would like to derive the analytical expression for the power spectrum of curvature perturbations.
At the linear level, the curvature perturbation $\zeta$ and the isocurvature modes $\sigma$ are defined as
\be
\delta\phi^a = \sqrt{2\epsilon}\zeta T^a + \sigma N^a
\ee
And their full equations of motion in terms of e-foldings are given by
\begin{align} \label{fulleq1}
\frac{d}{dN}\left(\frac{d\zeta}{dN} -\frac{\lambda}{\sqrt{2\epsilon}}\sigma\right) +(3-\epsilon+\eta)\left(\frac{d\zeta}{dN} -\frac{\lambda}{\sqrt{2\epsilon}}\sigma\right) + \frac{k^2}{a^2 H^2}\zeta & = 0, \\
\frac{d^2\sigma}{dN^2} +(3-\epsilon)\frac{d\sigma}{dN}  + \sqrt{2\epsilon}\lambda\left(\frac{d\zeta}{dN} -\frac{\lambda}{\sqrt{2\epsilon}}\sigma\right)+\frac{k^2}{a^2 H^2}\sigma + \frac{\mu^2}{H^2}\sigma & = 0~, \label{fulleq2}
\end{align}
where $\mu^2$ is the entropy mass of the isocurvature perturbations given by
\begin{equation}
\mu^2 \equiv V_{NN} + \epsilon H^2 \mathbb{R} + 3\Omega^2~.
\end{equation}
Thus besides $V_{NN}$, the turning effects and the curvature of the field manifold also contribute to the entropy mass. But in multi-field $\alpha$-attractors here, as shown in Figure~\ref{fig:bglambdagamma}, $\mu^2$ is mainly controlled by the $V_{NN}$ term. Then by \eqref{vnn2} and  the solution of $\vp$ in \eqref{Nphi}, we get
\begin{equation}
 \frac{\mu^2}{H^2} \approx  \frac{V_{NN}}{H^2} \approx 3B e^{-\sqrt{2}\vp} \approx \frac{3}{N } ,
\end{equation}
which provides a good analytical approximation as shown in Figure~\ref{fig:bglambdagamma}.

The exact solutions of the full equations \eqref{fulleq1} and \eqref{fulleq2} can be obtained only through numerical method, as we have shown in Figure~\ref{fig:power}.
But notice that the leading effect here comes from the coupled evolution of curvature and isocurvature modes after horizon-exit.
Thus for the analytical approximations, we can focus on super-horizon scales. There the isocurvature equation of motion reduces to
\begin{equation}
 3\frac{d\sigma}{dN}+ \frac{\mu^2}{H^2}\sigma \simeq 0 .
\end{equation}
If we focus on the mode that exits horizon at $N_*$ with amplitude $\sigma_*$, then we get the following solution for its evolution
\begin{equation}
 \sigma (N)  \approx \sigma_\ast \frac{N}{N_\ast}~.
\end{equation}
Remember that e-folding number is counted backwards from the end of inflation. Thus this solution shows the decay of the isocurvature perturbation outside of the horizon.
The evolution of the normalized isocurvature power spectrum $P_\mathcal{S}$ is shown in Figure~ \ref{fig:power}, where $\mathcal{S}=\sigma/\sqrt{2\epsilon}$. As we see, the analytical approximation above successfully captures the super-horizon decay, compared with the numerical result.

Next, we look at the curvature perturbation, which after horizon-exit is sourced by the the isocurvature modes in the following way
\begin{equation}
\frac{d \zeta}{dN} = \frac{\lambda}{\sqrt{2\epsilon}}\sigma .
 \label{eqn:superhorizonR}
\end{equation}
Also for the mode exits horizon at $N_*$ with amplitude $\zeta_*$, we get the solution
\begin{equation}
 \zeta(N) = \zeta_\ast + \int_{N_\ast}^{N} dN^\prime \ \frac{\lambda}{\sqrt{2\epsilon}}\sigma(N^\prime).
\end{equation}
As we noticed in \eqref{eqn:largephilambda}, $\lambda/\sqrt{2\epsilon}$ is nearly constant, thus it can be seen as unchanged after horizon-exit  $\lambda/\sqrt{2\epsilon}= \lambda_*/\sqrt{2\epsilon_*}$. Meanwhile, notice that since $\sigma$ is almost massless in the large-$\vp$ regime, one has the relation ${\sqrt{2\epsilon_*}}\zeta_\ast\simeq {\sigma_\ast}\simeq H/(2\pi)$.
Then the evolution of $\zeta$ is given by
\be
 \zeta(N)  =\zeta_\ast +  \frac{\lambda_\ast}{2} \frac{N^2-N_*^2}{N_*} \zeta_\ast .
\ee
These two contributions are uncorrelated with each other, since they come from the different parts of the quantized fluctuations.
Thus finally we can write down the power spectrum at the end of inflation ($N=0$)
\begin{align}
 P_{\zeta} = P_{\zeta}^{(1)} + P_{\zeta}^{(2)} & =  \frac{H^2}{4\pi^2}\frac{1}{2\epsilon_\ast}\left(1+ \frac{\lambda_\ast^2 N_*^2}{4}\right)   = \frac{H^2}{8\pi^2\epsilon_\ast} \left(1+ \gamma^2 \right) =\frac{H^2}{8\pi^2\epsilon_{\vp\ast}} ,
\end{align}
where we used the relation \eqref{epsgamma}, the expression of $\lambda$ \eqref{eqn:largephilambda}, and $\epsilon_\vp\simeq1/(4N^2)$. Therefore, we recover the same result as we got in $\delta N$ calculation.

It is interesting to note that, although the turning  effects play an important role in the intermediate calculation, they vanish in the final answer. There are two effects on the curvature perturbation in multi-field $\alpha$-attractors: first, the growth on super-horizon scales gives an enhancement factor  $\left(1+ \gamma^2 \right)$; secondly,  due to the motion in the $\theta$ direction, the slow-roll parameter $\epsilon$ is also enhanced by the same amount.
Thus as a consequence, these two changes cancel each other, and the final power spectrum of curvature perturbation here becomes the same as the single field result.

\end{appendices}

\bibliographystyle{JHEP}
\bibliography{bibfile}

\end{document}